\documentclass{article}
\title{Attribute-based Encryption for Attribute-based Authentication, Authorization, Storage, and Transmission in Distributed Storage Systems}
\author{Aubrey Alston (ada2145@columbia.edu)}

\usepackage[left=2.5cm,top=3cm,right=2.5cm,bottom=3cm,bindingoffset=0.5cm]{geometry}

\usepackage{fancyhdr}
\date{}

\usepackage{amsmath}
\usepackage{amssymb}
\usepackage{MnSymbol}
\usepackage{graphicx}
\usepackage{float}
\usepackage{amsthm}
\usepackage{algorithm}
\usepackage{algpseudocode}
\usepackage{algorithmicx}
\usepackage{program}
\usepackage{caption}

\theoremstyle{definition}
\newtheorem{definition}{Definition}[section]

\newcommand\blfootnote[1]{%
  \begingroup
  \renewcommand\thefootnote{}\footnote{#1}%
  \addtocounter{footnote}{-1}%
  \endgroup
}

\begin{document}

\maketitle

\section{Overview}

Attribute-based encryption \blfootnote{This work is the project report corresponding to an undergraduate research 
project completed with the Columbia University IDS Lab during the spring of 2016.} is a form of encryption which offers the capacity to encrypt data 
such that it is only accessible to individuals holding a satisfactory configuration of attributes.  
As cloud and distributed computing become more pervasive in both private and public spheres, 
attribute-based encryption holds potential to address the issue of achieving secure authentication, 
authorization, and transmission in these environments where performance must scale with security while 
also supporting fine-grained access control among a massively large number of consumers.  
With this work, we offer an example generic configurable stateless protocol for secure attribute-based 
authentication, authorization, storage, and transmission in distributed storage systems 
based upon ciphertext-policy attribute-based encryption (CP-ABE), discuss the experience of implementing 
a distributed storage system around this protocol, and present future avenues of work enabled by 
such a protocol.  The key contribution of this work is an illustration of a means by which any CP-ABE system may be 
utilized in a black-box manner for attribute-based authentication and cryptographically enforced 
attribute-based access control in distributed storage systems.
\newline\newline

\tableofcontents

\section{Background}

In this section, we (1) provide an introduction to attribute-based encryption, 
(2) briefly introduce the premise and significance of distributed storage, and 
(3) motivate a need for a secure attribute-based authentication, authorization, 
storage, and transmission protocol by identifying security concerns in the 
distributed systems underlying cloud environments.

\subsection{Attribute-based Encryption}

Attribute-based encryption (ABE) is a form of encryption first proposed by Sahai 
and Waters as an application of fuzzy identity-based encryption, a type of encryption
through which data is encrypted on the basis of individual identity as represented by a set 
of attributes \cite{WatersIBE}.  Goyal et. al. later explored the topic in more detail, presenting 
attribute-based encryption as a generalization of this construction in which data is 
encrypted according to some logical expression of attributes, called an access policy, 
such that encrypted data can only be decrypted if that policy is satisfied \cite{WatersABE}.
The authors of \cite{WatersABE} further defined a formal security setting for an ABE scheme:
in addition to the cryptographic guarantees (e.g. IND-CPA security) expected of encryption schemes in 
general, a useful ABE scheme must also be resistant to collusion attacks, defined as an attack 
in which one or more key holders attempt to decrypt data for which they are otherwise not authorized
by sharing keys.  
\break\break
In its discussion of attribute-based encryption, the authors of \cite{WatersABE} discuss
two forms of attribute-based encryption: ciphertext-policy ABE (CP-ABE) and key-policy ABE (KP-ABE).  
The core difference between ciphertext-policy ABE and key-policy ABE lies in the logical 
location of the access policy:  in CP-ABE, a set of attributes is associated with a key, and 
encrypted data can be decrypted only if those attributes satisfy the access policy associated 
with it; in KP-ABE, a set of attributes is associated with encrypted data, and data can be 
decrypted only if those attributes satisfy a policy associated with the key being used.  
To illustrate both the utility of ABE and the differences between CP-ABE and KP-ABE, 
consider the following two illustrative scenarios:
\break\break
\textit{Scenario 1}:  Company \textit{C} maintains a file \textit{F} which contains confidential salary information 
for employees in New York.  \textit{F} should be encrypted and stored such that it can only be read by
either the company CEO or managers in New York.

This end could be achieved using CP-ABE as follows:  first, grant users keys containing
attributes corresponding to their position and their office location; next, encrypt \textit{F} using
the policy ‘CEO or (Manager and New York).’  In the future, depending on the security of 
the chosen ABE scheme and its application context, only an individual whose key satisfies this 
policy may decrypt and read the contents of \textit{F}.  
\break
\break
\textit{Scenario 2}: Agency \textit{D} maintains a file \textit{F} which contains a report on confidential material.  Each
section of the report contains content surrounding some subset of topics.  An individual
\textit{A} is authorized to read any classified material so long as it concerns both Europe and 
agriculture.

This end could be achieved using KP-ABE as follows: first, encrypt \textit{F} in segments,
associating a set of attributes corresponding to the set of sensitive topics contained.
Next, issue \textit{A} a key with the policy ‘Europe and Agriculture.’  \textit{A} will then, depending on the
relative security of the ABE scheme, be able to decrypt and read only those sections
containing information on both Europe and agriculture.
\break
\break
As is iterated in \cite{WatersABE}, CP-ABE and KP-ABE are suited towards different use cases.  
Intuitively, between the two components required for decryption--ciphertext and key--
the one with which the policy is associated imposes a more selective condition than the other.  
As we are concerned with use cases in which access to remote data by consumers is controlled 
by policy, this work makes exclusive use of CP-ABE.

\subsection{Ciphertext-policy Attribute-based Encryption}
A valid CP-ABE scheme must provide four algorithms: \textit{Setup}, \textit{Encrypt},
\textit{GenerateKey}, and \textit{Decrypt} \cite{Waters08CPABE}, defined in greater detail below:

\begin{itemize}
\item \(Setup(\lambda, U)\): Takes a security parameter \(\lambda\) and a description of attributes
within the system \textit{U} and outputs a set of public parameters \textit{PK} and a master key
\textit{MK}.

\item \(Encrypt(PK, M, A)\): Takes a set of public parameters \textit{PK}, a message to encrypt \textit{M},
and an access policy \textit{A} expressed in terms of attributes and outputs a ciphertext \textit{C}, 
generally assumed to also contain A.

\item \(GenerateKey(MK, S)\): Takes a master key \textit{MK} and a set of attributes that should be
associated with the key \textit{S} and outputs a private key \textit{SK}.

\item \(Decrypt(PK, C, SK)\): Takes a set of public parameters \textit{PK}, a ciphertext to decrypt \textit{C}, 
  and a secret key \textit{SK} and outputs the decrypted message \textit{M}.
\end{itemize}

\break
\break
In works such as \cite{MultiABE}, authors have since introduced extensions to CP-ABE which 
allow for the construction of multi-authority CP-ABE schemes.  In multi-authority and 
hierarchical CP-ABE schemes, multiple authorities work with one another to generate 
keys (at times with each authority being responsible for some possibly disjoint 
subset of attributes).  

Although these variants of ABE define slightly different interfaces for each algorithm, for the sake of this work, 
any reference to \textit{Setup} or \textit{GenerateKey} may be assumed to reference the interface provided by any one individual 
CP-ABE scheme (single-authority, multiple-authority, or otherwise) so long as said scheme 
(a) defines \textit{Encrypt} and \textit{Decrypt} as specified and (b) does not assume a static universe of attributes.

\subsection{Cloud Computing and Distributed Storage}

Cloud computing is a model which enables on-demand access to a shared pool of configurable 
computing resources; a primary objective of cloud computing is the capacity to rapidly 
provision and release these resources with minimal effort on the part of the service 
provider as required by the measured needs of consumers; this allows applications 
to arbitrarily scale to meet the needs of one consumer or billions of consumers \cite{NISTCloud}.  
NIST defines three service models of cloud computing: Software as a Service (SaaS), 
Platform as a Service (PaaS), and Infrastructure as a Service (IaaS): SaaS allows 
shared access to the provider’s application running on cloud infrastructure; PaaS 
provides shared access and management of languages, tools, and services required to 
deploy an application; IaaS provides shared access to fundamental hardware resources 
such as disk space, networks, and CPU cores \cite{NISTCloud}.

Data-intensive applications that ultimately either rely on or support SaaS or PaaS service 
models inherently require logical data storage and access at scale. While all service models 
guarantee that storage space may be provisioned as necessary, support for shared distributed 
storage among these resources is required by applications which need to operate concurrently 
on contiguous data.  To meet this need in a manner optimal for their respective use cases, 
various distributed storage systems have been developed.  To provide a non-exhaustive set of 
examples: GFS is a distributed storage system developed by Google to provide general-case 
fault-tolerant, scalable storage over commodity hardware \cite{GFS}; HDFS is a distributed 
storage system implemented to meet the storage needs of a specific cluster computing 
framework \cite{HDFS}; DynamoDB is a distributed storage system implemented by Amazon 
to support the access pattern defined by its shopping cart feature \cite{Dynamo}.  

\subsection{Security in Distributed Storage Systems}

While the many distributed storage systems which exist are intentionally highly performant, 
available, and scalable, their initial design does not directly guarantee congruent 
security of stored data.  In this section, we demonstrate the deficiencies that exist
in such systems with respect to confidentiality and integrity of data stored, and we
further illustrate the limits of common approaches to addressing these deficiencies.
\break\break
\textit{Confidentiality}.  Consider the confidentiality of data stored by a distributed storage system. 
If, as described in \cite{GFS},\cite{HDFS}, and \cite{Dynamo}, stored resources are served to any requesting entity, 
the confidentiality of data intended to be accessible by a specific subset of users cannot be guaranteed. 
A distributed storage system expected to ensure confidentiality of data must implement some means of access control;
to do so, such a system must also have a means of reliably discerning an authorized user from a non-authorized user
(potentially via authentication) and subsequently enforcing access control.  Confidentiality, however, is still 
not absolutely guaranteed by sound authentication, authorization, and access control: if data is stored in the clear, 
any unauthorized party gaining access to a storage node inevitably gains access to the data residing at that node.  
Similarly, if data or control messages are transmitted in the clear, an unauthorized party may gain knowledge of 
data directly or indirectly by simply eavesdropping. 
\break\break
\textit{Integrity}.  Consider the integrity of data stored by a distributed storage system.  
If both system components and consumers accept control messages and data transferred blindly, 
an attacker could potentially influence either the server or the client to 
accept invalid data by modifying the stream of data sent.  Similarly, if an attacker gains access to a storage node, 
integrity is trivially compromised by means of the fact that such an attacker may then arbitrarily modify data.
This latter risk to integrity is somewhat offset by the structure of distributed storage systems making use of
replication: in such systems, breaches to integrity may be corrected by means of copying from a replica.
\break\break
Though offering secure access control, storage, and transmission may allow a distributed storage system to
preserve the confidentiality and integrity of stored data, the question of how a distributed storage system
might do so in a genuinely secure manner without affecting performance or scalability of the system remains a 
challenge.  In practice, systems attempt to resolve these concerns by means of either symmetric-key
encryption and/or public-key encryption; however, the constraints imposed by conventional symmetric and public-key 
cryptography degrade feasibility and security at scale in an environment in which resources are not restricted based upon 
the unique identity of the consumer.

As an example, HDFS, a popular distributed storage system used commonly in support
of cluster computing, implements an approach which utilizes symmetric-key encryption: HDFS introduces
the notion of an encryption zone, in which the aim is for files to be securely shared between consumers
in a zone \cite{HDFSEncryption}.  Each file in an encryption zone is encrypted using a symmetric cipher using
a random key, which is itself stored encrypted under a zone encryption key, shared
among all consumers in the zone \cite{HDFSEncryption}.  This approach, however, merits caution with respect
to both security and feasibility.  With respect to security, this approach may not be optimal, as it 
requires keys to be shared between a large number of consumers for large zones; with respect to feasibility, 
this approach may not be optimal, as each consumer must store one key for each to which he or she has access.  

Unlike symmetric-key and group-based approaches to addressing the above issues, an approach utilizing 
attribute-based encryption inherently comes with the capacity to express fine-grained access control 
without having to track the identity of users or perform burdensome key management at a scale linear 
in the number of groups of files to which a user has access.  Furthermore, attribute-based cryptography
admits a means for allowing consumers to authenticate and be authorized on a basis other than identity,
a pattern more suited for distributed storage systems seeking security at scale.

\subsection{Related Work}

Attribute-based encryption has been identified previously as a potentially useful basis for security 
in distributed storage systems and cloud computing.  Early works on attribute-based encryption 
such as \cite{WatersABE} and \cite{WatersIBE} reference applications for which ABE may be promising, 
including broadcast log encryption, an application also essential for the auditing of the 
distributed systems underlying cloud environments.  

Other works, such as \cite{Horvath2015}, \cite{AnonyControl}, and \cite{HierarchicalABE}, 
point out potential gaps in ABE as given canonically, aiming specifically to provide some extension of 
ABE which might satisfy the requirements of a storage application in a cloud environment.  
\cite{HierarchicalABE} aims to introduce a provably secure hierarchical ABE (H-ABE) construction to 
allow for more flexibility in system architectures employing ABE; \cite{AnonyControl} aims 
to provide a multi-authority ABE construction to fulfill a similar need while also fulfilling anonymity 
requirements; \cite{Horvath2015} identifies a particular necessary capability of large-scale 
operating systems--key revocation--and provides a provably secure construction meant to reduce the 
computational burden of key revocation at the cost of a slight overhead in encryption and decryption.   CP-ABE, Wang et. al. 
have proposed a system called Sieve which utilizes KP-ABE in untrusted cloud environments, with a particular 
focus on web services. \cite{Sieve}.

This work is distinct from previous works in that, rather than providing a specific ABE construction or a system to be
utilized by other systems and services, 
it provides (a) a generic configurable protocol which uses CP-ABE to achieve secure access, storage, 
and transmission of distributed resources in generic distributed storage systems and (b) a generic threat model 
that may be used to analyze any such protocol providing access control on the basis of attributes.  
As a result, a system utilizing this protocol may opt to choose one from among all ABE schemes according to its 
unique performance and security expectations; likewise, the threat model provided may be used as a basis of 
analysis for alternative systems and protocols.  The protocol proposed further separates the function of 
recovery from the cryptographic scheme used, providing a feasible means of directly performing key
revocation and recovery without having to rely on ad-hoc qualities of the configured cryptosystem.

\section{The Protocol}

In this section, we (1) introduce a general, high-level architecture abstraction used 
to represent the system which will house our proposed protocol, (2) briefly introduce 
preliminaries of our proposed protocol, and finally (3) outline the protocol.

\subsection{System Setting}

The basic protocol introduced is defined with respect to a system which provides access to 
distributed resources to consumers (semi-anonymous or identified) who interact with the system 
on the basis of their attributes.  Such a system is assumed to be composed of some number of system components
called \textit{nodes}, each of which taking one of the following roles with respect to the protocol: 
\begin{itemize}
\item \textit{Authorization Node}: Entity responsible for performing authorization of consumers on some
subset of attributes recognized within the system.
\item \textit{Service Node}: Entity responsible for providing a secure interface to some subset of 
partitions of distributed resources.  Consumers communicate with service nodes to access or modify 
distributed resources.
\item \textit{Authority}: Entity responsible for creating keys and arbitrating the global universe
of attributes recognized by the system.  An authority may delegate to other components to perform
key distribution\footnote{The protocol given admits potentially many key distribution schemes.  
See Future Work}.
\end{itemize}

\begin{figure}
\centerline{\includegraphics[scale=0.85]{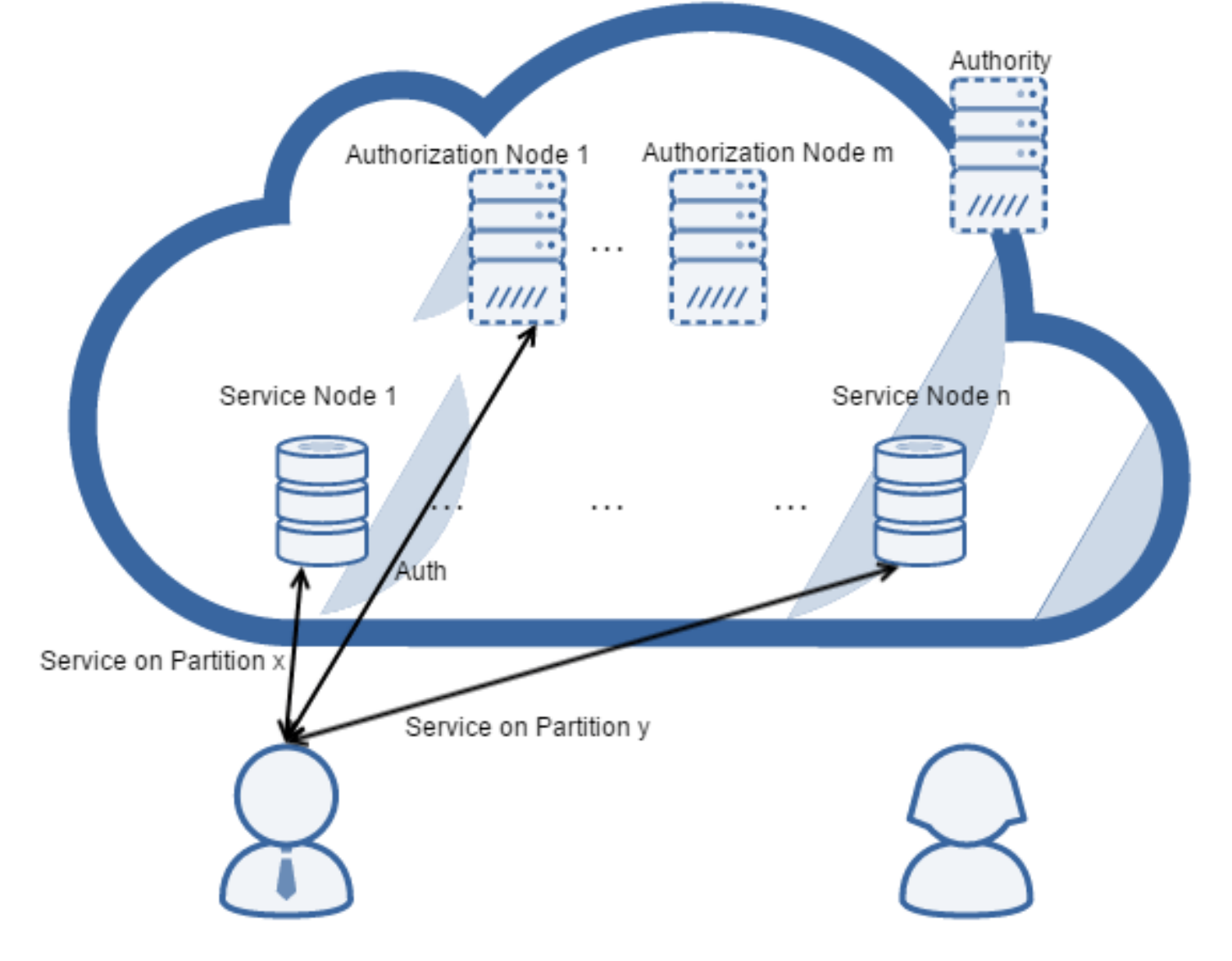}}
\caption{Example configuration of roles among system components.}
\label{fig:one}
\end{figure}

\break 

Figure 1 provides an illustration of such an architecture containing \textit{m} authorization nodes 
and \textit{n} disjoint service nodes.  A consumer first authenticates to and is authorized by an authorization 
node for some subset of attributes, and he or she is then serviced by service node 1 on partition \textit{x} and 
service node \textit{n} on partition \textit{y}.  This specific configuration of component roles
reflects a system architecture using many (\textit{m}) masters and many (\textit{n}) storage nodes.  

Though the example in Figure 1 is specific, the abstract system setting described may be 
configured to map to any general distributed storage system: this configuration may be
modified, for example, to map to a single-master system by reducing the number of authorization nodes to one;
alternatively, it could be modified to map to quorum-like system architectures by making the sets of
authorization nodes and service nodes non-disjoint\footnote{This latter case
would require the coordination techniques applied in quorum-like systems to also be applied to authorization nodes.  See Future Work}.  
Note also that this setting allows a system to decouple the location of consumer authorization 
from the location of the master key holder (the authority) within the utilized ABE scheme; 
any such coordination of roles may be easily extended to employ multiple or hierarchical 
authorities if using a multiple-authority ABE scheme.

\subsection{Preliminaries}

The proposed protocol makes use of additional abstractions defined with respect to attributes.

\begin{definition}{\textit{Representation}}
The representation of an attribute is the functional component of a named attribute within an ABE construction.
\footnote{Example: in a hypothetical ABE scheme in which an attribute is manipulated or applied mathematically using
a group element unique to that attribute, the representation of an attribute would be its respective group element.}
\end{definition}

\begin{definition}{\textit{Volatile Attribute}}
A volatile attribute is an attribute whose representation may be expected to change.
\end{definition}

\begin{definition}{\textit{Manifest Attribute}}
A manifest attribute is an attribute whose representation must necessarily be known to parties outside of the
set of system components.
\end{definition}

\begin{definition}{\textit{Bound Attribute}}
A bound attribute is an attribute whose representation is coupled with the state of some resource.
\end{definition}

To illustrate the meanings and uses of these abstractions, consider the task of encrypting a large file using ABE.  
If the policy on the file involves the attribute \textit{A}, \textit{A} becomes a bound attribute.  
If \textit{F} must be accessible to external consumers, \textit{A} also becomes a manifest attribute.  
We would hope that \textit{A} is not also a volatile attribute, as any change in its representation would 
cause the system to need to re-encrypt \textit{F}; similarly, if a volatile attribute \textit{A} were to become manifest, 
it may become necessary to propagate the new representation of \textit{A} to interested consumers.

In addition to these abstractions of attributes, the protocol proposed defines additional constructs
to be used in the authentication and authorization process. 

\begin{definition}{\textit{Master Session Token (MST)}}
An object produced as a result of authorization which authenticates consumer attributes to a service node
and contains authorization from an authorization node.
\end{definition}

\begin{definition}{\textit{Validity Attribute}}
A validity attribute is a volatile, manifest attribute assigned to consumers whose primary role is to attest to the validity
of the key of a requesting consumer.  When a consumer authenticates to an authorization node, the consumer must specify a set of
held validity attributes, and the produced MST is sent such that it may only be used if the consumer holds all of the attributes
with which he or she authenticated as well as the validity attributes specified.  A validity attribute will never be bound to a 
file resource served by the protocol.
\end{definition}

\subsection{Protocol Definition}

The proposed protocol is composed of seven configured algorithmic components:

\begin{enumerate}
\item An attribute-based encryption scheme \(E_{ABE}\)
\item An attribute-based encryption scheme under a variable-length mode of encryption \(E_{CHAIN}\)\footnote{This may be derived from \(E_{ABE}\).  See Implementation and Future Work.}
\item (Optional) An attribute-based signature scheme \(S_{SIG}\)
\item A public-key signature scheme \(S_{PSIG}\)
\item A variable-length symmetric-key encryption scheme \(E_{SYM}\)
\item A MAC scheme \(MAC\)
\item A secure key exchange routine \(R_{KE}\)
\end{enumerate}

The only constraint placed on these components is that \(E_{ABE}\) must be able to encrypt ciphertexts at least
as large as the size of keys and parameters for \(E_{SYM}\).  If there is no single entity that is to be absolutely
identifiable among system components, \(S_{SIG}\) should be configured as an attribute-based signature scheme.  If 
\(S_{SIG}\) is not configured, public-key \(S_{PSIG}\) is used in its place.

In addition to the configured algorithms, the following parameters must also be configured:
\begin{enumerate}
\item Role assignment for system components (mapping system components among authorities, service nodes, and authorization nodes)
\item Parameters \(x\) and \(u\), respectively configuring the number of validity attributes to be used by
the system and the minimum number of validity attributes a consumer must advertise in order to authenticate.
\end{enumerate}

\textit{System initialization}.  At system initialization, the configured authority (or authorities) perform the
\(Setup\) algorithm as specified by \(E_{ABE}\), with an initial universe \(U\) containing an attribute \textit{SN}
to be issued for provisioned service nodes, an attribute \(a_i\) for each initial generic manifest attribute 
to associate with consumers and to be bound to stored resources, an attribute \(AN_{A_j}\) for each subset of
attributes \(A_j\) to be authorized exclusively by some subset of authorization nodes, and all validity attributes 
\(v_1 . . . v_x\).

\textit{Addition of new components}.  When a new authorization node is provisioned, that node is issued a 
ABE key containing the attribute \(AN_{A_j}\) corresponding to the subset of attributes \(A_j\) for which 
it is responsible for authorizing and a public key pair if \(S_{SIG}\) has not been configured.  
When a new service node is provisioned, that node is issued an ABE key containing the 
attribute \(AN_{A_j}\) and all validity attributes.
\break\break
Note: This protocol allows component locations and responsibilities to be public knowledge.  For the sake of simplicity
and generality, this work assumes that there is a public index of all public parameters (including a whitelist of 
locations, responsibilities, and public parameters of \(E_{ABE}\)) that reflects changes immediately; 
private changes (including the issuance of keys) are assumed to be communicated by means of a generic secure 
key distribution mechanism.  Exploration of specific key distribution and knowledge propagation 
mechanisms for specific applications of this protocol may be valuable; see Future Work.
\break\break
The protocol defines four routines: \textit{Attribute-Authenticate}, \textit{Put}, \textit{Get}, and \textit{Write}.
\subsubsection{\textit{Attribute-Authenticate}}  A consumer \textit{C} negotiates with an authorization node \textit{M} to
authenticate a set of attributes \textit{A} it would like to use during its session.  At the end of a valid 
negotiation, \textit{C} receives a \textit{MST} from \textit{M}.
\begin{enumerate}
\item \textit{C} and \textit{M} perform \(R_{KE}\).  All subsequent communication is encrypted using \(E_{SYM}\) and authenticated
 using \(MAC\) under the derived key.
\item \textit{C} sends \(A = { a_1, . . ., a_n } \mid\mid V = { v_1, . . ., v_u } \mid\mid  TTL_{req}  \mid\mid  PK_{self} \) to \textit{M}.  

\(TTL_{req}\) is a requested TTL for the \(MST\); \(PK_{self}\) is a public key for the requesting consumer (not tracked by the system).
\item \textit{M} derives the subset of \textit{A}, \(A'\), which it is allowed to authorize.
\item \textit{M} chooses \textit{expiry} as the minimum of \(TTL_{req}\) and a maximum TTL plus the current time.
\item \textit{M} generates two random \(E_{SYM}\) key-parameter pairs \(K_1\), \(K_2\) and a random \(R\).
\item \textit{M} computes and sends \(K' = Encrypt_{E_{ABE}}(PK, K_1, (\wedge^n_{i=0}a_i) \wedge (\wedge^u_{j=0}v_j))\) to \textit{C}.
\item \textit{M} computes \(MST_1 = E_{ABE}(K_2, SN \wedge (\wedge^u_{j=0}v_j))  \mid\mid  A'  \mid\mid  expiry  \mid\mid  R  \mid\mid  PK_{self}\).
\item \textit{M} computes \(MST_2 = S_{SIG}(MST_1, M_{private})\)

\(M_{private}\) corresponds to either the private
component of the authorizaton node's public-key keypair or the authorizaton node's ABE private key, depending on the
configuration of the protocol.
\item \textit{M} computes \(MST_3 = Encrypt_{E_{SYM}}(K_1  \mid\mid  expiry  \mid\mid  R, K_2)\).
\item \textit{M} computes \(MST = MST_1  \mid\mid  MST_2  \mid\mid  MST_3  \mid\mid  M_{public}\).

\(M_{public}\) is the public key for the authorization node if \(S_{SIG}\) is configured as a public-key
signature scheme.  Otherwise, \(M_{public}\) is an indication of the authorization node attribute
\(AN_{A_j}\) held by the authorization node.
\item \textit{M} computes and sends \(MST' = Encrypt_{E_{SYM}}(MST,K_1)\) to \textit{C}.  
\item \textit{C} decrypts \(K'\) to obtain \(K_1\) and retrieves
\(MST\) by decrypting \(MST'\) with \(K_1\).
\end{enumerate}
When a service node \textit{N} receives a \(MST\), it must verify the validity of the \(MST\).  A service node
\textit{N} verifies a \(MST\) as follows:
\begin{enumerate}
\item Validate that the contained \(M_{public}\) corresponds to a valid authorization node for the set of 
attributes \(A'\) specified by the MST.  If \(S_{SIG}\) is configured as an attribute-based signature 
scheme, \textit{N} simply checks that the specified \(M_{public}\) contains \(AN_{A_j}\) corresponding
to the subset of attributes specified.
\item \textit{N} validates the content of the MST via the signature.
\item \textit{N} obtains \(K_2\) from \(MST_1\) and decrypts \(MST_3\) using \(K_2\).  
\item \textit{N} then verifies that \textit{R} and \textit{expiry} agree between \(MST_1\) and \(MST_3\). 
\item \textit{N} verifies that the MST has not expired.
\end{enumerate}
\subsubsection{\textit{Put}} A consumer \textit{C} negotiates with a service node \textit{N} to persist
a new resource in the system.
\begin{enumerate}
\item \textit{C} and \textit{N} perform \(R_{KE}\).  All subsequent communication is encrypted using \(E_{SYM}\) and authenticated
 using \(MAC\) under the derived key.
\item \textit{C} sends \textit{N} a MST.
\item \textit{C} sends \textit{N} an identifier for the new resource, an access policy for new resource, 
and the size of new resource to be transferred. 
\item \textit{N} validates the MST and further validates that the attributes authorized in the 
MST satisfy the policy specified.
\item \textit{C} sends encrypted resource content \textit{D} as \(Encrypt_{E_{SYM}}(Encrypt_{E_{CHAIN}}(D, policy), K_1)\), 
optionally through an encryption proxy, and \(S_{PSIG}(Encrypt_{E_{CHAIN}}(D, policy), SK_{self})\).  \(SK_{self}\) is
the private component of the \(PK_{self}\) included in the sent MST. 
\item \textit{N} accepts if and only if the signature verifies.
 \end{enumerate}
\subsubsection{\textit{Get}} A consumer \textit{C} negotiates with a service node \textit{N} to request a distributed resource \textit{F}.
\begin{enumerate}
\item \textit{C} and \textit{N} perform \(R_{KE}\).  All subsequent communication is encrypted using \(E_{SYM}\) and authenticated
 using \(MAC\) under the derived key.
\item \textit{C} sends \textit{N} an identifier for \textit{F} and any application-related request parameters.
\item \textit{C} sends \textit{N} a MST.
\item \textit{N} validates the MST and further validates that the attributes authorized in the MST satisfy the policy on
\textit{F}.
\item \textit{N} sends \(Encrypt_{E_{SYM}}(Encrypt_{E_{CHAIN}}(F, policy), K_1)\) to \textit{C}.
\item \textit{C} decrypts using \(K_1\) and its private ABE key.
\end{enumerate}
\subsubsection{\textit{Write}} A consumer \textit{C} negotiates with a service node \textit{N} to modify a distributed resource \textit{F}.
\begin{enumerate}
\item \textit{C} and \textit{N} perform \(R_{KE}\).  All subsequent communication is encrypted using \(E_{SYM}\) and authenticated
 using \(MAC\) under the derived key.
\item \textit{C} sends \textit{N} an identifier for \textit{F} and any application-related request parameters (e.g. range to write).
\item \textit{C} sends \textit{N} a MST.
\item \textit{N} validates the MST and further validates that the attributes authorized in the MST satisfy the policy on \textit{F}.
\item \textit{C} sends modified resource data \(D'\) as \(Encrypt_{E_{SYM}}(Encrypt{E_{CHAIN}}(D', policy)), K_1)\), 
optionally through an encryption proxy, and \(S_{PSIG}(Encrypt_{E_{CHAIN}}(D', policy), P_{SK})\).
\item \textit{N} accepts if and only if the signature matches.
\end{enumerate}

\section{Analysis}

In this section we introduce a general threat model for a distributed storage system utilizing 
attributes for authentication, authorization, transmission, and storage and discuss 
the security of the proposed protocol within the framework of this threat model.

\subsection{Threat Model}

In a distributed storage environment consisting of \textit{consumers}, \textit{authorization nodes}, \textit{service nodes}, 
and \textit{authorities} interacting on a basis of attributes, we express a threat model with respect to 
\textit{assets}, \textit{agents}, and \textit{adversaries}.

\begin{definition}{\textit{Asset}}
An asset is any resource accessed, modified, or used via the system, including private keys.
\end{definition}

\begin{definition}{\textit{Agent}}
An agent is an entity which either actively or passively interacts with the system.
\end{definition}

\begin{definition}{\textit{Adversary}}
An adversary is an agent who attempts to induce compromise in the system.
\end{definition}

Our threat model qualifies security as a distributed storage system's ability to avoid,
prevent, and react to the compromise of concerned assets.  An asset is considered \textit{compromised} if
any of the following conditions are met:
\begin{itemize}
\item An agent not granted a satisfying set of attributes with respect to an implicit or explicit
access policy on an asset gains access, degrading confidentiality of the concerned asset.
\item An agent not granted a satisfying set of attributes causes a permanent modification, 
degrading integrity of the concerned asset.
\item An agent granted a satisfying set of attributes is deceived into accepting
an asset that has been tampered with, degrading both availability and integrity.
\end{itemize}

Our model further classifies real and potential compromises according to properties of their
consequences and severity.
\begin{definition}{\textit{Online-recoverable compromise}}
A compromise is online-recoverable under a protocol if the system may feasibly harden to
correct and prevent said compromise without halting system operation.
\end{definition}
\begin{definition}{\textit{Local compromise}}
A compromise is local under a protocol if it may only affect the asset it concerns.
\end{definition}
\begin{definition}{\textit{Forward compromise}}
A compromise is forward under a protocol if it cannot induce additional compromise of resources
transferred or held in the past.
\end{definition}

Our threat model assumes that an adversary has the ability to eavesdrop and tamper 
with any communication between agents and that an adversary has substantial computational resources; 
adversaries may collude and share resources with one another, including compromised assets.  
We define a weaker definition of this threat model in which the attack surface is restricted 
solely to the routines and communication defined by the protocol; we define a stronger 
definition where the attack surface is assumed to be expanded to subsume individual system 
components, client machines, operators, and network resources, allowing for direct exfiltration 
of assets as well as insider adversaries and impersonators.

\subsection{Analysis of Proposed Protocol}

In this section, we analyze the given protocol through the lens of the given threat model
defined in the previous section.  For each attack surface exposed by the protocol, we 
enumerate and classify risks of compromise conditioned upon the configuration of 
the protocol.

\subsubsection{\textit{Attribute-Authenticate}}
Under the weak threat model, we demonstrate that the Attribute-Authenticate routine is secure with respect to all 
transferred assets against active and passive attackers, conditioned upon the 
security guarantees and the collusion resistance of the configured \(E_{ABE}\), 
the chosen plaintext security guarantees of the configured \(E_{SYM}\), the 
guarantees of the configured \(R_{KE}\), and the forgery resistance of the configured \(MAC\).

Assume first the capabilities of a passive adversary under the weaker definition of the threat model.  
In the very first step of the routine, the consumer and the authorization node perform \(R_{KE}\), and 
so an eavesdropping adversary is not able to learn anything about the shared key that is used during the routine 
so long as the guarantees of \(R_{KE}\) hold.  If an adversary modifies messages sent as a part of \(R_{KE}\), 
he or she induces only a temporary disagreement in the shared key (renegotiation can be performed again), 
which will be immediately detected by the consumer and authorization node upon integrity verification using \(MAC\)
so long as \(MAC\) is forgery resistant.  Thus, conditioned on the guarantees of the configured 
\(R_KE\) and the forgery resistance guarantees of \(MAC\), there is no risk of compromise of the shared 
key under the defined threat model.

Because all subsequent communication is encrypted using the derived shared key and authenticated via \(MAC\), 
no eavesdropping adversary is able to gain any information about messages sent so long as \(E_{SYM}\) is IND-CPA secure, 
any modification to message content may be detected by either party so long as \(MAC\) is forgery resistant, and 
corrupted messages may be re-sent and/or renegotiated so long as a communication channel remains open between consumer and 
authorization node.  Conditioned on the chosen plaintext security of the configured \(E_{SYM}\) and the forgery resistance 
guarantees of \(MAC\), no asset exchanged during \textit{Attribute-Authenticate} has a risk of compromise by an active or passive
adversary under the weak definition of our threat model.

With regard to \textit{Attribute-Authenticate}, an active participating adversary would attempt to induce a future compromise by obtaining a 
MST which authorizes him or her for attributes he or she does not have.  Depending on the guarantees of the configured 
\(E_{ABE}\), however, an adversary should not be able to obtain the plaintext of the MST because it is encrypted (under \(E_{SYM}\)) 
using a key which is encrypted (under \(E_{ABE}\)) using an access policy that is only satisfied if the consumer holds all of the 
attributes advertised.  Thus, under the inherent decryption guarantees and the chosen plaintext security of \(E_{ABE}\), 
no adversary may obtain or distinguish a MST for which he or she should not be authorized under the weak threat model.  It is also not possible, 
by means of the collusion resistance of the configured \(E_{ABE}\) scheme, that adversarial key holders 
collude to produce a satisfying \(E_{ABE}\) key.  As the only asset transferred by \textit{Attribute-Authenticate}, the MST, 
is thus secure against both passive and active attackers, we conclude that the attack surface exposed by \textit{Attribute-Authenticate}
is secure under the weak threat model.
\break\break
\textit{Strong threat model.}  Under the strong threat model, we demonstrate that compromise of any asset concerned by 
the Attribute-Authenticate routine is at the very least forward and online-recoverable.

Consider the capabilities of a worst-case adversary who has gained compromised access to the \(E_{ABE}\) private key, 
the \(S_{SIG}\) key-pair, and the \(S_{PSIG}\) key-pair during the \textit{Attribute-Authenticate} routine. 
As a result of \textit{Attribute-Authenticate} utilizing \(R_{KE}\) 
without intermediate storage of shared keys, the adversary is unable to re-derive or gain knowledge of non-compromised 
messages exchanged in the past, conditioned on the guarantees and assumptions of \(R_{KE}\); 
thus, any compromise of the assets held by the authorization node is necessarily a forward compromise.  
Note that compromise of the \(S_{SIG}\) key-pair of a valid authorization node allows an adversary to forge master 
session tokens, potentially compromising assets exchanged in the other routines of the protocol, 
meaning that such a compromise is not local.  Assuming that some number of system components and at least one authority 
remain uncompromised, such a compromise is yet online-recoverable, as the protocol admits at least one means 
of online recovery:
\begin{enumerate}
\item Revoke the volatile authorization node attribute bound to forged MSTs by changing its representation
and propagating it to all other active system components.  This will cause verification of forged MSTs to fail immediately
if the protocol uses an attribute-based signature scheme.  
\item Blacklist the compromised authorization node and any associated public keys.  This will cause verification of the
forged MSTs to fail in the case that they are signed using \(S_{PSIG}\).
\item Provision a new authorization node having a replacement key for the revoked attribute.
\item Propagate knowledge of the existence of the new authorization node.
\end{enumerate}

Consider now the capabilities of a worst-case adversary who has gained compromised access to the private ABE 
key held by a valid consumer.  Under the \textit{Attribute-Authenticate} routine, such an adversary may arbitrarily request, 
receive, and use MSTs he or she should not have.  Because illegally obtained MSTs do not allow an adversary 
to gain knowledge of assets transferred in the past, this compromise is a forward compromise.
This compromise is yet online-recoverable, as the protocol admits at least one key revocation mechanism that prevents
use of this compromised key:
\begin{enumerate}
\item Revoke at least one of the manifest volatile validation attributes associated with the compromised key by changing
its representation, presenting knowledge of the new representation publicly.  Keys using these validation attributes are thus 
effectively revoked, as they will not be serviced under the protocol; granularity of revocation is inherently
configured by the number of validation attributes used within the system.
\item Re-issue keys with which the re-issued validation attribute is associated.
\end{enumerate} 

\subsubsection{\textit{Put}}  Under the weak threat model, we demonstrate that the \textit{Put} routine is secure 
with respect to all transferred assets against active and passive attackers depending on the chosen plaintext 
security guarantees of the configured \(E_{ABE}\), the chosen plaintext security guarantees of the configured \(E_{SYM}\), 
the guarantees of the configured \(R_{KE}\), the forgery resistance of the configured \(MAC\), \(S_{SIG}\), and \(S_{PSIG}\).

In the case of a passive third party adversary, for the same reasons as given in our analysis of \textit{Attribute-Authenticate}, 
conditioned on the guarantees of the configured \(R_{KE}\) and the collision-resistance guarantees of \(MAC\), 
there is no risk of compromise of the shared key under the defined threat model.  As in the case of \textit{Attribute-Authenticate}, 
because all subsequent communication is encrypted using the derived key and authenticated using \(MAC\), 
no eavesdropping adversary is able to gain any information about messages sent so long as \(E_{SYM}\) is IND-CPA secure, 
any modification to message content by an active third party may be detected by either participating party so long as \(MAC\) is forgery resistant, 
and corrupted messages may be re-sent or renegotiated so long as a communication channel remains open between consumer and service node; 
thus, no asset exchanged during \textit{Put} has risk of compromise by an active or passive third party.

With regard to \textit{Put}, a participating adversary may attempt to directly compromise a distributed resource asset by 
inducing a service node to commit a new resource for which he or she should not be authorized.  
In order to do so, however, such an adversary would need to present a valid MST.  Under the weak threat 
model, we have shown that such an adversary cannot obtain a usable MST since he or she does not hold the 
required attributes.  In order to coerce the service node to accept a resource in this setting, 
the adversary would need to forge a MST, which should not be possible, conditioned upon the assumptions and guarantees
of the configured \(S_{SIG}\) and \(S_{PSIG}\).  For these reasons, \textit{Put} is secure from active participating adversaries
under the weak threat model.
\break\break
\textit{Strong threat model.}  Under the strong threat model, we demonstrate that compromise of any asset concerned by the \textit{Put} 
routine is at the very least forward and online-recoverable.  

Consider the capabilities of a worst-case adversary who has gained compromised complete access to a service node, 
including its \(E_{ABE}\) key and stored encrypted data.  Within the context of \textit{Put}, such an adversary may 
either (a) reject or deny service to users, (b) allow spurious writes to data, (c) passively comply with the protocol, 
further exfiltrating and compromising future versions of the stored encrypted data, or (d) modify stored data.  
By the nature of this adversary, all data compromised in this scenario is forward compromised: past versions of data or 
transferred assets are not stored; in addition to this, compromised data is encrypted and yields information only for an 
individual which holds a satisfying ABE key.  Assuming that some number of system components and at least one authority 
remain uncompromised, such a compromise is online-recoverable, as the protocol admits at least one means 
of online recovery:
\begin{enumerate}
\item Revoke the volatile service node attribute\footnote{This protocol may be easily adapted to allow for the system
to make use of multiple service node attributes as opposed to one.} held by the compromised service node.
\item Blacklist the compromised service node and any associated public keys.
\item Provision a new service node having a replacement key for the revoked attribute.
\item Propagate replacement service keys to active service nodes.
\end{enumerate}

Consider now the capabilities of a worst-case adversary who has gained compromised 
access to the private ABE key, the signature key-pair, or and the MST held by a valid consumer.  
Under the \textit{Put} routine, data to be written must be signed using \(S_{PSIG}\) and the private 
key corresponding to \(PK_{self}\).  Assume first that this adversary has only the MST of a valid consumer.  
Unless this adversary also has the ability to forge a signature corresponding to the public key \(S_{PSIG}\),
the service node will reject the put, meaning that control of a MST alone is merely a local compromise, 
as it cannot induce further compromises without more information, conditioned on the assumptions and guarantees of
the configured \(S_{PSIG}\).  If the adversary does have control 
of the consumer \(S_{PSIG}\) private key, such an adversary would likely also have control of the consumer private \(E_{ABE}\) key.  
In either case, this adversary would be able to write arbitrarily through the service node.   
As was the case for \textit{Attribute-Authenticate}, this forward compromise is also online-recoverable, as the protocol 
admits at least one key revocation mechanism which will invalidate both compromised keys and MSTs.
If ever only the \(S_{PSIG}\) private key is compromised along with the MST, the protocol is self-recovering, 
as the MST will eventually expire. 

\subsubsection{\textit{Get}}  Under the weak threat model, we demonstrate that the \textit{Get} routine is secure 
with respect to all transferred assets against active and passive attackers depending on the chosen plaintext 
security guarantees of the configured \(E_{ABE}\), the chosen plaintext security guarantees of the configured \(E_{SYM}\), 
the guarantees of the configured \(R_{KE}\), the forgery resistance of the configured \(MAC\), \(S_{SIG}\), and \(S_{PSIG}\).

In the case of a passive third party, for the same reasons as given for \textit{Attribute-Authenticate}, 
conditioned on the guarantees of the configured \(R_{KE}\) and the forgery resistance guarantees of \(MAC\), 
there is no risk of compromise of the shared key under the defined threat model.  
As in the case of \textit{Attribute-Authenticate}, because all subsequent communication is encrypted using the derived key 
and verified using \(MAC\), no eavesdropping adversary is able to gain any information about messages sent 
so long as \(E_{SYM}\) is IND-CPA secure, any modification to message content by an active eavesdropper may 
be detected by either party so long as \(MAC\) is resistant to forgeries, and corrupted messages may be re-sent 
or renegotiated so long as a communication channel remains open between consumer and service node; 
no asset exchanged during \textit{Get} has risk of compromise by an active or passive third party.

With regard to \textit{Get}, an actively participating adversary may attempt to directly compromise a distributed resource asset 
by inducing a service node to send it data for which it is not authorized.  In order for the service node 
to transfer anything, however, such an adversary would need to present a valid MST.  Under the weak threat 
model, we have shown that such an adversary cannot obtain a usable MST since he or she does not hold the required 
attributes.  In order to coerce the service node to accept a modification in this setting, the adversary would need to 
forge a MST, which should not be possible given the guarantees of \(S_{PSIG}\).  In addition to this, any data which is 
transferred is encrypted under a policy, and so the adversary would not be able to gain anything from what is 
transferred unless he or she has also compromised a satisfying \(E_{ABE}\) key.
\break\break
\textit{Strong Threat Model}.  Under the strong threat model, we demonstrate that compromise of 
any asset concerned by the Get routine is at the very least forward and online-recoverable. 

Consider the capabilities of a worst-case adversary who has gained compromised complete access to a service node, 
including its \(E_{ABE}\) key and stored encrypted data.  Within the context of \textit{Get}, such an adversary may send 
requesting consumers corrupt data as desired, exfiltrate encrypted data as desired, or deny service as desired.  
By the nature of this adversary, all compromises of data in this 
scenario is forward compromised: past versions of data or transferred assets are not stored.  Among information 
gained would be the MSTs of all valid consumers; however, these have no use without further compromise of the consumer 
\(S_{PSIG}\) private keys.  Assuming that some number of system components and at least one authority remain uncompromised, 
such a compromise is online-recoverable, by means of, for example, the recovery mechanism described in the case of \textit{Put}.

Consider now the capabilities of an adversary who has gained compromised access to either the private ABE key, 
the signature key-pair, or a MST held by a valid consumer.  Under the \textit{Get} routine, resources 
requested must be signed using \(S_{PSIG}\) and the private key corresponding to \(PK_{self}\).  By the same argument as 
for \textit{Put}, only the scenario in which the adversary has further compromised a satisfying ABE key, the consumer private 
\(S_{SIG}\) key, and the consumer MST yields a compromise; this compromise is a forward, online-recoverable compromise which may be 
corrected via key revocation or simple MST expiry as described in previous sections.

\subsubsection{\textit{Write}}  Under the weak threat model, we demonstrate that the \textit{Write} routine is secure 
with respect to all transferred assets against active and passive attackers depending on the chosen plaintext 
security guarantees of the configured \(E_{ABE}\), the chosen plaintext security guarantees of the configured \(E_{SYM}\), 
the guarantees of the configured \(R_{KE}\), the forgery resistance of the configured \(MAC\), \(S_{SIG}\), and \(S_{PSIG}\).

In the case of a passive third party adversary, for the same reasons as given in our analysis of \textit{Attribute-Authenticate}, 
conditioned on the guarantees of the configured \(R_{KE}\) and the collision-resistance guarantees of \(MAC\), 
there is no risk of compromise of the shared key under the defined threat model.  As in the case of \textit{Attribute-Authenticate}, 
because all subsequent communication is encrypted using the derived key and authenticated using \(MAC\), 
no eavesdropping adversary is able to gain any information about messages sent so long as \(E_{SYM}\) is IND-CPA secure, 
any modification to message content by an active third party may be detected by either participating party so long as \(MAC\) is forgery resistant, 
and corrupted messages may be re-sent or renegotiated so long as a communication channel remains open between consumer and service node; 
thus, no asset exchanged during \textit{Write} has risk of compromise by an active or passive third party.

With regard to \textit{Write}, a participating adversary may attempt to directly compromise a distributed resource asset by 
inducing a service node to commit a new resource for which he or she should not be authorized.  
In order to do so, however, such an adversary would need to present a valid MST.  Under the weak threat 
model, we have shown that such an adversary cannot obtain a usable MST since he or she does not hold the 
required attributes.  In order to coerce the service node to accept a resource in this setting, 
the adversary would need to forge a MST, which should not be possible, conditioned upon the assumptions and guarantees
of the configured \(S_{SIG}\) and \(S_{PSIG}\).  For these reasons, \textit{Write} is secure from active participating adversaries
under the weak threat model.

\textit{Strong Threat Model}.  Under the strong threat model, it is the case that compromise of any asset 
concerned by the \textit{Write} routine is at the very least forward and online-recoverable by the same argument as 
that for the forward, online-recoverability of compromises concerned by \textit{Put}.

\subsection{Scalable Security}

A major requirement for any scalable storage system is that security is capable of scaling with performance.  To analyze
the scalability of our protocol, we put forth a framework for quantifying the scalable security of a protocol for a scalable
system and further analyze our proposed protocol through this framework.
\break\break
\textit{A measure for scalable security.}  We define scalable security as the capacity for a system to maintain its
instantaneous security guarantees as it scales up or down arbitrarily.  To measure the support provided by a protocol 
for achieving this quality, we introduce a construct which we call \textit{security-preserving scaling effort}.
\begin{definition}{\textit{Security-preserving Scaling Effort}}
In a distributed system composed of categories of components \(C_1, C_2, . . ., C_n\) and a scaling action \(A\), 
the security-preserving scaling effort of \(A\) is a function \(S_A( \mid\mid C_1 \mid\mid , . . .,  \mid\mid C_n \mid\mid )\) which provides a reasonable asymptotic 
upper bound on the number of operations required such that the security guarantees of the system 
before \(A\) are the same as those after \(A\) completes. 
\end{definition}
In our analysis of the security-preserving scaling effort of our protocol, we define categories of components 
\(A, AN, SN, C\), respectively representing authorities, authorization nodes, service nodes and consumers.  Our analysis
focuses on a set of scaling actions restricted to adding or removing an authority, adding or removing an authorization node,
adding or removing a service node, and adding or removing a consumer; our scaling effort is defined with respect to the operations of 
sending messages to a single system component and performing keying operations.
\break\break
In this section, we demonstrate that our protocol requires a scaling effort linear in the number of authorities
for all defined scaling actions except for key revocation, which itself is linear per-revoked key in the number of authorities and 
authorization nodes.
\break\break
\textit{Adding or removing an authority}.  Our protocol does not involve direct interaction with authorities.  With respect to 
effort required in adding or removing an authority, the effort required by our protocol is bounded by the number of operations
required to make active authorities aware of the change, making the change public to some interface to the key distribution 
mechanism, and potentially deriving a new key for the new authority provisioned.
\begin{equation}
S_A \in O( \mid\mid A \mid\mid ) 
\end{equation}

\textit{Adding or removing an authorization node}.  Within our protocol, when an authorization node is added or removed, it must 
potentially be granted both an ABE key and a public keypair.  Because the configured ABE scheme may be multi-authority, it is 
possible that messages will have to be sent from each authority.  Our protocol further assumes that knowledge of the locations
of authorization nodes be made public; this may be achieved in practice by making the change known to some set of public indexes.  
\begin{equation}
S_A \in O( \mid\mid A \mid\mid )
\end{equation}

\textit{Adding or removing a service node}.  Within our protocol, when a service node is added or removed, it must 
potentially be granted an ABE key.  Because the configured ABE scheme may be multi-authority, it is 
possible that messages will have to be sent from each authority.  Our protocol further assumes that knowledge of the locations
of service nodes be made public; this may be achieved in practice by making the change known to some set of public indexes.   
\begin{equation}
S_A \in O( \mid\mid A \mid\mid )
\end{equation}

\textit{Adding a consumer}.  As long as a consumer has a public keypair and a valid ABE key, a consumer may interact with
the protocol.  Thus, to add a new consumer, the only effort required is the generation of a new ABE key the sending of that 
key to the consumer.  In the worst case, the configured ABE scheme is multi-authority, and so messages must be sent from each authority
to the new consumer.
\begin{equation}
S_A \in O( \mid\mid A \mid\mid ) 
\end{equation}

\textit{Removing a consumer}.  In order to remove a consumer, that consumer's ABE key must be invalidated.  In order to do this,
the representation of at least one validity attribute \(v_x\) must be changed, thereby revoking the keys of all consumers \(C \in v \) requiring 
\(v_x\) to properly authenticate.  Keys must be re-issued to all of such consumers.  In order for the change to be enforced, the 
change must be propagated to all active authorization nodes.
\begin{equation}
S_A \in O( \mid\mid v \mid\mid  \mid\mid A \mid\mid  +  \mid\mid AN \mid\mid ) 
\end{equation}

\section{Future Work}

\subsection{Implementation}

To explore the feasibility and utility of this protocol, we have undertaken the 
implementation\footnote{At the time of this work, the partial
implementation of this system may be found here: https://github.com/ad-alston/aefs}
of a simple manifestation of this protocol.

The target system is a simple single-master, many-worker distributed file system in the tone of GFS\cite{GFS} 
with a flat namespace and simple access control achieved through an implementation of the protocol described in this 
work.  The specific configuration of the protocol implemented is as follows:
\begin{enumerate}
\item Waters08\cite{Waters08CPABE} as a CP-ABE scheme.
\item A custom extension of Waters08 using AES-256 to achieve variable-length encryption using CP-ABE.
\item RSA-2048 as a public-key signature scheme.
\item AES-256/CBC-MAC for variable-length symmetric-key encryption with \(MAC\)
\item Diffie-Hellman as a key exchange routine.
\end{enumerate}

This and future implementations of the proposed protocol will be valuable so as to demonstrate
the practical challenges and factors to consider in evaluating different configurations of 
the given protocol.

\subsection{Performance Analysis in Context}

The introduction of the devised scheme will undoubtedly affect the performance of the system which uses it.  
Future work should concern itself with applying the given protocol to existing systems and performing an a
nalysis in performance relative to increased security.

\subsection{Improvement and Standards for CP-ABE}

Though the proposed protocol provides a means of addressing the issue of scalable authorization, 
authentication, and transmission in distributed storage systems, its design, analysis, and 
implementation suggest areas of exploration relating to both feasibility and utility of 
CP-ABE schemes.  Current schemes are non-standard, varying vastly in performance and complexity;
likewise, the lack of maturity of ABE schemes has prevented the appearance of different modes of
encryption under CP-ABE schemes.  A specific direction for improvement would be the introduction
of variable-length encryption modes for CP-ABE, modes desired by the protocol proposed in this work.

\subsection{Logging and Compromise Detection}

The proposed protocol admits methods for recovering from compromise both internally and externally.  
When master session tokens or the private keys of users are used to interact with the system, the 
mechanics of the protocol have a side effect that they generate information at service and authorization 
nodes about the keys being used in the form of validation attributes.  Given some volume of the access 
patterns for both volatile validation attributes and manifest non-volatile attributes of users, how can 
the logging and analysis of these access patterns provide additional benefit to the system employing the protocol?

One possible area of exploration involves the detection of and reaction to the compromise of private keys 
and tokens among users.  It may be possible to issue validation attributes on some identifiable basis in
order to be able to detect anomalies in key usage.  As a simple example, the system may distribute 
validation attributes on the basis of geographic region.  If there is, say, a high frequency of authorization 
requests from a region in which a certain validation attribute was not issued, the system may be able to 
infer that keys within that region have been compromised and recover accordingly.  Another venue for future work 
could be the exploration of using the structure and function of validation attributes
to trace both traitors and eager attackers: by introducing the equivalent of honey-attributes, it may be possible to 
detect attempted compromise and/or attempted treason by otherwise valid consumers.

\subsection{Coordination and Replication}

Many distributed storage systems such as DynamoDB\cite{Dynamo} opt for non-centralized architectures that rely 
heavily on coordination and quorum-like techniques.  With respect to the mechanics of the protocol, how might 
the coordination techniques used by these systems be transitioned to allow for the implementation of 
mechanisms which make use of coordination among authorization nodes and service nodes?  Similarly, 
what benefits can be offered by these mechanisms?

A direct application of such coordination lies in internal system auditing in systems which employ replication.  
If a system component is suspicious to an individual user (or other system component), it may be useful for service 
nodes to vote on the integrity of the state or actions of that component; similarly, if attacks affecting integrity 
can be identified, coordination among replicas may be used to recover.

\subsection{Key Distribution, Knowledge Propagation, and Recovery}

Though the analysis of the protocol under the strong threat model defined in this paper identifies 
that recoveries are online-recoverable, the protocol itself does not directly concern itself with how a 
system should be adapted to perform online recovery.  In the case of key revocation, future work should 
explore the mechanics of system components responsible for coordinating with the system authority(ies), propagating 
attribute representation changes, and delivering renewed keys as needed.  For recovery in the case of compromised system components, 
future work should also explore means for a system to provision replacements and segregate those components which have been compromised.

Recent work by Wang et. al. introduces a method of using key homomorphism and secret sharing to achieve a means of
key revocation and two-factor authentication \cite{Sieve}.  While they propose using this method to reconstruct 
the master secret key under the used KP-ABE scheme, it may be possible within the scope of our protocol to use this method to 
devise a secure key distribution mechanism to be used to reconstruct master secret keys among controlled 
system components before re-distributing keys.  Similarly, key homomorphism may be used within the scope of our protocol
as a measure for re-encrypting compromised stored assets without system components gaining knowledge of the asset's
plaintext.

\section{Conclusion}

In this work, we have introduced a protocol for attribute-based encryption for authentication, authorization, storage, 
and transmission of distributed resources which addresses the problem of achieving scalable security in distributed 
storage systems that require fine-grained access control among large numbers of semi-anonymous consumers. We further 
analyzed this protocol through a general threat model which can be used to analyze the security guarantees of a 
distributed storage system under these constraints.  Through design, implementation, and analysis of this protocol, 
we have identified future opportunities for exploration admitted by this protocol which additionally allow for 
internal and external threat detection, automatic recovery, and automated key revocation.

% Start of "Sample References" section
\bibliographystyle{acm}
\bibliography{abeprotocol}

\begin{thebibliography}{10}

\bibitem{HDFSEncryption}
{\sc Apache}.
\newblock Transparent encryption in hdfs.
\newblock
  https://hadoop.apache.org/docs/r2.7.2/hadoop-project-dist/hadoop-hdfs/TransparentEncryption.html/,
  2016.

\bibitem{MultiABE}
{\sc Chase, M., and Chow, S.~S.}
\newblock Improving privacy and security in multi-authority attribute-based
  encryption.
\newblock In {\em Proceedings of the 16th ACM Conference on Computer and
  Communications Security\/} (New York, NY, USA, 2009), CCS '09, ACM,
  pp.~121--130.

\bibitem{Dynamo}
{\sc DeCandia, G., Hastorun, D., Jampani, M., Kakulapati, G., Lakshman, A.,
  Pilchin, A., Sivasubramanian, S., Vosshall, P., and Vogels, W.}
\newblock Dynamo: Amazon's highly available key-value store.
\newblock {\em SIGOPS Oper. Syst. Rev. 41}, 6 (Oct. 2007), 205--220.

\bibitem{GFS}
{\sc Ghemawat, S., Gobioff, H., and Leung, S.-T.}
\newblock The google file system.
\newblock In {\em Proceedings of the Nineteenth ACM Symposium on Operating
  Systems Principles\/} (New York, NY, USA, 2003), SOSP '03, ACM, pp.~29--43.

\bibitem{WatersABE}
{\sc Goyal, V., Pandey, O., Sahai, A., and Waters, B.}
\newblock Attribute-based encryption for fine-grained access control of
  encrypted data.
\newblock In {\em Proceedings of the 13th ACM Conference on Computer and
  Communications Security\/} (New York, NY, USA, 2006), CCS '06, ACM,
  pp.~89--98.

\bibitem{Horvath2015}
{\sc Horv{\'a}th, M.}
\newblock {\em SOFSEM 2015: Theory and Practice of Computer Science: 41st
  International Conference on Current Trends in Theory and Practice of Computer
  Science, Pec pod Sn{\v{e}}{\v{z}}kou, Czech Republic, January 24-29, 2015.
  Proceedings}.
\newblock Springer Berlin Heidelberg, Berlin, Heidelberg, 2015,
  ch.~Attribute-Based Encryption Optimized for Cloud Computing, pp.~566--577.

\bibitem{AnonyControl}
{\sc Jung, T., Li, X.-Y., Wan, Z., and Wan, M.}
\newblock Control cloud data access privilege and anonymity with fully
  anonymous attribute-based encryption.
\newblock {\em IEEE Transactions on Information Forensics and Security 10}, 1
  (Dec. 2014), 190--199.

\bibitem{NISTCloud}
{\sc Mell, P.~M., and Grance, T.}
\newblock Sp 800-145. the nist definition of cloud computing.
\newblock Tech. rep., Gaithersburg, MD, United States, 2011.

\bibitem{WatersIBE}
{\sc Sahai, A., and Waters, B.}
\newblock Fuzzy identity-based encryption.
\newblock In {\em Proceedings of the 24th Annual International Conference on
  Theory and Applications of Cryptographic Techniques\/} (Berlin, Heidelberg,
  2005), EUROCRYPT'05, Springer-Verlag, pp.~457--473.

\bibitem{HDFS}
{\sc Shvachko, K., Kuang, H., Radia, S., and Chansler, R.}
\newblock The hadoop distributed file system.
\newblock In {\em Proceedings of the 2010 IEEE 26th Symposium on Mass Storage
  Systems and Technologies (MSST)\/} (Washington, DC, USA, 2010), MSST '10,
  IEEE Computer Society, pp.~1--10.

\bibitem{Sieve}
{\sc Wang, F., Mickens, J., Zeldovich, N., and Vaikuntanathan, V.}
\newblock Sieve: Cryptographically enforced access control for user data in
  untrusted clouds.
\newblock In {\em Proceedings of the 13th Usenix Conference on Networked
  Systems Design and Implementation\/} (Berkeley, CA, USA, 2016), NSDI'16,
  USENIX Association, pp.~611--626.

\bibitem{HierarchicalABE}
{\sc Wang, G., Liu, Q., and Wu, J.}
\newblock Hierarchical attribute-based encryption for fine-grained access
  control in cloud storage services.
\newblock In {\em Proceedings of the 17th ACM Conference on Computer and
  Communications Security\/} (New York, NY, USA, 2010), CCS '10, ACM,
  pp.~735--737.

\bibitem{Waters08CPABE}
{\sc Waters, B.}
\newblock Ciphertext-policy attribute-based encryption: An expressive,
  efficient, and provably secure realization.
\newblock In {\em Proceedings of the 14th International Conference on Practice
  and Theory in Public Key Cryptography Conference on Public Key
  Cryptography\/} (Berlin, Heidelberg, 2011), PKC'11, Springer-Verlag,
  pp.~53--70.

\end{thebibliography}
                             % Sample .bib file with references that match those in
                             % the 'Specifications Document (V1.5)' as well containing
                             % 'legacy' bibs and bibs with 'alternate codings'.
                             % Gerry Murray - March 2012

%\section{Typical references in new ACM Reference Format}
%\cite{Abril07}

% Acknowledgments
%\begin{acks}
%The authors would like to thank Dr. Maura Turolla of Telecom
%Italia for providing specifications about the application scenario.
%\end{acks}

% Bibliography

% History dates

% Electronic Appendix
% \elecappendix

\end{document}